\documentclass[10pt,fleqn]{article}

\usepackage{amssymb}

\newcommand{\name}[1]{\begin{flushleft}
                       \LARGE \bf #1
                       \end{flushleft}\vspace{-3mm}}

\newcommand{\Author}[1]{\begin{flushleft}
                       \it #1 \end{flushleft}}

\newcommand{\Adress}[1]{\begin{flushleft}
                       \it #1 \end{flushleft}}

\newcommand{\AbsEng}[1]{
    \begin{flushright}
    \begin{minipage}{120mm}
     \small   #1
    \end{minipage}
    \end{flushright}
}

\newcommand{\be}{\begin{equation}}
\newcommand{\ee}{\end{equation}}
\newcommand{\ba}{\hspace*{-5pt}\begin{array}}
\newcommand{\ea}{\end{array}}
\newcommand{\p}{\partial}

\newcommand{\pbf}[1]{\mbox{\mathversion{bold}$#1$}}

\begin{document}

\name{On a possible approach\\ to the variable-mass problem}

\medskip

\noindent{published in {\it Nuclear Physics B}, 1968, {\bf 7}, P. 79--82.}

\Author{Wilhelm I. FUSHCHYCH and Ivan Yu. KRIVSKY}

\Adress{Institute of Mathematics of the National Academy of
Sciences of Ukraine,\\ 3 Tereshchenkivska  Street, 01601 Kyiv-4,
UKRAINE}

\noindent {\tt URL:
http://www.imath.kiev.ua/\~{}appmath/wif.html\\ E-mail:
symmetry@imath.kiev.ua}

\AbsEng{The mass operator $M$  is introduced as an independent dynamical variable
 which is taken as the translation generator $P_4$ of the inhomogenous De Sitter group.
The classification of representations of the algebra $P(1,4)$
of this group is performed and the corresponding $P(1,4)$
invariant equations for variable-mass particles are written out. In this way we have succeeded,
in particular, in uniting the ``external'' and ``internal'' $(SU_2)$
symmetries in a non-trivial fashion.}

\medskip

\noindent
The idea of variable mass has been considered by many authors in connection with
the mass-spectrum and unstable-particles problems (see e.g. refs.~[1,~2]).
Since for the stable free particle $M_0^2=P_0^2-{\pbf P}^2$,
it is supposed that the square of variable-mass operator (in the presence of interaction,
of course) is defined by $M^2\equiv P_0^2-{\pbf P}^2$.

In connection with the problems mentioned, the idea seems to be attractive to consider
the rest mass as a variable $M$, on the same footing with the three-momen\-tum~${\pbf P}$.
It is natural to realize this idea in such a way that we define the mass operator~$M$
as an independent dynamical variable $P_4$
like the components of three-momentum~${\pbf P}$.
By this the correspondence principle with the fixed-rest-mass theory demands the
free operators of energy, three-momentum and variable mass to satisfy the condition
\be
P_0^2 ={\pbf P}^2+M^2\equiv {\pbf P}^2+P^2_4
\ee
in this case too. It is obvious that such a definition of the mass operator is
more general than the above and is non-trivial (see below) even in the case of absence
of interaction.

The presence of a dynamical variable $M\equiv P_4$ in the $p$-representation
in addition to the three-momentum ${\pbf P}$
makes it inevitable to introduce in the $x$-representation (in quantum mechanics)
an additional dynamical variable besides the three coordinates~${\pbf x}$.
It is natural to take for this the canonical conjugate of $M$,
which will be denoted below as $\tau\equiv x_4$.
It should be noted in this context that at least the corresponding principle with the
fixed-rest-mass theory (not to speak of deeper physical arguments) does not allow to
consider the time $t\equiv x_0$ as a dynamical variable (e.g. as canonically conjugated
to $P_0$) in the variable mass theory too. Further, if we do not want to violate the conventional
connection between the momentum and configuration spaces, the variable-mass concept
discussed here requires to study the group of transformations which conserve
the five-dimensional form $x^2\equiv t^2-{\pbf x}^2 -\tau^2\equiv x_\mu^2$,
$\mu=0,1,2,3,4$, i.e.,  the inhomogenous De Sitter group, the algebra of which we call
$P(1,4)$, in analogy to the algebra $P(1,3)$ of the Poincar\'e group.

To write $P(1,4)$-invariant equations for free particles with variable mass,
we make use of the classification of the irreducible representations of $P(1,4)$.
Analogously to the case of the algebra $P(1,3)$ (ref.~[3]),
we consider four classes of representations which correspond to the values $P^2=\kappa^2>0$,
$P^2=0$ and $P^2=-\eta^2<0$ of the invariant
\be
P^2=P_0^2-{\pbf P}^2-P_4^2 \equiv P_\mu^2, \qquad \mu=0,1,2,3,4.
\ee

The algebra $P(1,4)$ which is determined by the generators $P_\mu$ and $M_{\mu\nu}$
has, besides eq.(2), two other invariants:
\be
V\equiv -\frac 14 M_{\mu\nu} w_{\mu\nu}, \qquad W\equiv \frac 12 w_{\mu\nu}^2,
\ee
where
\be
w_{\mu\nu}\equiv -\frac 12 \varepsilon_{\mu\nu \alpha\beta \gamma}M_{\alpha\beta} P_\gamma.
\ee

For class I (when $P^2=\kappa^2>0$) , in the system ${\pbf P}=P_4=0$, we have
\be
\hat S^2 \equiv P_0^{-2} W+2 P_0^{-1}V =({\pbf M}+{\pbf R})^2 =s(s+1)\hat 1,
\ee
\be
\hat I^2 \equiv P_0^{-2} W-2P_0^{-1} V =({\pbf M}-{\pbf R})^2=I(I+1)\hat 1,
\ee
where $s, I=0,\frac 12,1,\ldots$ and
\be
{\pbf M}\equiv (M_{23}, M_{31}, M_{12}), \qquad {\pbf R}\equiv (M_{14}, M_{24}, M_{34}).
\ee

It follows from eqs.(5) and (6) that in the case I all the representations of $P(1,4)$
are unitarity and finite-dimensional (with respect to $s$ and $I$) and are labelled by
two numbers $s$ and $I$. These symbols are naturally to be identified with spin and
isospin of the free particle with variable mass $m$, and, owing to $p_0^2\geq \kappa^2$,
one can understand the parameter $\kappa$  (which is the boundary value of energy) as a
``bare rest mass'' of this particle.

The $P(1,4)$-invariant (in the Foldy [4] sence) equation for the wave function
$\psi(x)\equiv \psi(t,{\pbf x}, x_4)$ of such a particle (and antiparticle) with arbitrary
$s$ and $I$  has the form
\be
\left( \beta \sqrt{{\pbf P}^2 +P_4^2 +x^2}-i\frac{\p}{\p t}\right) \psi_{s_3I_3}(t,{\pbf x},x_4)=0,
\qquad \beta=\left(\begin{array}{cc} \hat 1 & 0 \\
0 & -\hat 1\end{array} \right),
\ee
where $-s\leq s_3\leq s$, $-I\leq I_3\leq I$.

Thus, the variable-mass concept discussed has made it possible to unite non-trivially the
``external'' ($P(1,3)$) and ``internal'' ($SU_2$) symmetries which was not successfully
done by the conventional approach to the variable mass~[2].

For the class II (when $P^2=0$), in the system $P_1=P_2=P_4=0$ we have
\be
P_3^{-1} V =-{\pbf M}{\pbf P}', \qquad P_3^{-2}W ={\pbf P}'^2,
\ee
where $P'_i\equiv M_{0i}+M_{i4} P_0 P_4^{-1}$, $i=1,2,3$.

The generators ${\pbf P}'$  and ${\pbf M}$ are those of the algebra $P(3)$
which are evoked by the group of translations and rotations in 3-dimentional Euclidean space.
Since the spectrum of $W$ is continuous in this case, its values are, obviously,
difficult to interpret physically in an acceptable fashion. If we put $W=0$
when $V=0$ ven if ${\pbf M}\not= 0$; an additional invariant
\be
W'\equiv {\pbf M}^2 =s(s+1)\hat 1
\ee
appears, so that in this case all the representations are unitary and finite
dimensional and are labelled by the spin $s=0,\frac 12, 1,\ldots$.

The $P(1,4)$-invariant equation for the wave function of the particle (and an\-ti\-par\-tic\-le)
with variable mass $m$ and arbitrary spin $s$ has the form
\be
\left( \beta \sqrt{{\pbf P}^2 +P_4^2 } -i \frac{\p}{\p t}\right) \psi_{s_3} (t,{\pbf x}, x_4)=0.
\ee

For the class III (when $P^2=-\eta^2<0$),  in the system $P_0={\pbf P}=0$
we have
\be
V=\pm \eta {\pbf M}{\pbf N}, \qquad W=\eta^2 ({\pbf N}^2 -{\pbf M}^2),
\ee
where ${\pbf N}\equiv (M_{01}, M_{02}, M_{03})$. The generators
${\pbf M}$  and ${\pbf N}$ are those of the algebra $O(1,3)$
of the homogenous Lorentz group and $V$, $W$ given in~(12)
are its invariants. Therefore, according to ref.~[5],  in this case all the unitary
irreducible representation of $P(1,4)$ are infinite dimensional and are labelled by
the numbers $\eta$, $l_0$ and $l_1$, where $\eta$  is arbitrary real, $-1\leq l_1\leq 1$
if $l_0=0$ and $l_1$ is imaginary of $l_0=\frac 12, 1, \ldots$.

In this case the $P(1,4)$-invariant equation for the wave function has the form
\be
\left( \beta\sqrt{{\pbf P}^2 +P_4^2 -\eta^2} -i\frac{\p}{\p t}\right) \psi_{ll_3}^{l_0 l_1}
(t,{\pbf x},x_4)=0,
\ee
where $l-l_0=0,1,2,\ldots$, $-l\leq l_3\leq l$. We have got in this way, the infinite-dimensional
equation for the wave function. The physical sence of the numbers
$\eta$, $l_0$, $l_1$, $l_3$ is not as clear as in the former cases.

Note, by the way, that recently the infinite-dimensional equations have been in\-ten\-si\-ve\-ly
discussed~[6] (though the physical arguments underlying them are still rather poor).
Following the authors~[6], one can try to find some physical sense for the numbers
$\eta$, $l_0$, $l_1$, $l_3$. However, it seems that the cases I and II can more directly
be related to the problems of mass spectrum and unstable systems than the case III.

For the sake of completeness we mention that in the case IV when $P_0={\pbf P}=P_4=0$,
we have $V=W=0$. The generators $M_{\mu\nu}$
remained are those of the algebra $O(1,4)$ of the homogenous De Sitter group,
all representations of which are well known (see e.g. ref.~[7]) and the corresponding
equations are written out in ref.~[8].

Here we have written down the $P(1,4)$-invariant equations for variable mass free particles,
more exactly, for elementary systems with respect to $P(1,4)$
(but, of course, non-elementary with respect to $P(1,3)$).
The construction of the quantum field theory on the basis of these equations
and the introduction of interaction, in the framework of the Lagrange formalism,
are performed by total analogy with the conventional theory. However, as a first
step to the problems of mass spectrum and unstable systems it is expedient to
introduce some interactions in the quantum mechanical equations and to study
the corresponding models which is the subject of the next publications.

\newpage

\begin{enumerate}

\footnotesize

\item Mathews P.T., Salam A., {\it Phys. Rev.}, 1958, {\bf  112}, 283; \\
Lurcat F., Strongly decaying particles and relativistic invariance, Preprint, Orsay, 1968.

\item O'Raifeartaigh L., {\it Phys. Rev. Letters}, 1965, {\bf 14}, 575; \\
Fushchych W.I., {\it Ukrainian Phys. J.}, 1968, {\bf  13}, 362.

\item Wigner E.P., {\it Ann. Math.}, 1939, {\bf 40}, 149;\\
Shirokov Yu.M., {\it JETP (Sov. Phys.)}, 1957, {\bf  33}, 1196.

\item Foldy L., {\it Phys. Rev.}, 1956, {\bf  102},  568.

\item Gelfand I.M., Minlos R.A., Shapiro Z.Ya., Representations of the rotation and Lorentz
Groups and their applications, Moscow, Fizmathgiz, 1958 (in Russian).

\item Fronsdal C., {\it Phys. Rev.}, 1967, {\bf  156}, 1665; \\
Nambu Y., {\it Phys. Rev.}, 1967, {\bf 160}, 1171;\\
Takabayashi T., {\it Progr. Theor. Phys.}, 1967, {\bf 37},  767.

\item Newton T.D., {\it Ann. Math.}, 1950, {\bf  51}, 730;\\
Gelfand I.M., Naimark M.A., Unit\"are Darstellungen der klassischen Gruppen,
Akademie-Verlag, Berlin, 1957.

\item Sokolik G.A., Group methods in the theory of elementary particles,
Atomizdat, Moscow, 1965 (in Russian).
\end{enumerate}
\end{document}